\documentclass[floatfix,aps,prb,showpacs,twocolumn,latexsym]{revtex4}
\usepackage{amssymb}

\usepackage{latexsym}
\usepackage[dvips]{graphicx}
\usepackage{amsmath}

\input{tcilatex}

\begin{document}

\title{ Diffuse emission in the presence of an inhomogeneous spin-orbit
interaction for the purpose of spin filtration}
\author{A.~Shekhter,$\;$ M.~Khodas, and A.~M.~Finkel'stein}

\begin{abstract}
A lateral interface connecting two regions with different strengths of the
Bychkov-Rashba spin-orbit interaction can be used as a spin polarizer of
electrons in two dimensional semiconductor heterostructures [Khodas \emph{%
et al.}, Phys. Rev. Lett. \textbf{92}, 086602 (2004)]. In this paper
we consider the case when one of the two regions is ballistic, while
the other one is diffusive. We generalize the technique developed
for the solution of the problem of the diffuse emission to the case
of the spin-dependent scattering at the interface and determine the
distribution of electrons emitted from the diffusive region. It is
shown that the diffuse emission is an effective way to get electrons
propagating at small angles to the interface that are most
appropriate for the spin filtration and a subsequent spin
manipulation. Finally, a scheme is proposed of a spin filter device,
see Fig.~\ref{fig:diffuse-filter}, that creates two almost fully
spin-polarized beams of electrons.
\end{abstract}

\affiliation{Department of Condensed Matter Physics, Weizmann Institute of
Science,Rehovot, 76100, Israel}
\maketitle


\section{introduction}

Recently we proposed \cite{KSF2003} to use a lateral interface between two
regions of the two-dimensional (2D) electron gas with different strengths of
the Bychkov-Rashba \cite{BychkovRashba84} spin-orbit (SO) interaction as a
spin-polarizing element for the purposes of spintronics. \cite%
{DattaDas1990,Wolf2001} The lateral interface introduces the
space-varying SO interaction which leads to spin-dependent
refraction of spin carriers passing through the
interface.\cite{lateralinterface} Consequently, an electron beam
with a nonzero angle of incidence after passing through the
interface splits into beams with different spin polarizations
propagating in different directions (see Fig.~\ref{fig:refraction}
for the scattering at the lateral interface between the regions with
and without SO interaction). Using an interface with an
inhomogeneous SO interaction as a principal element of spin-based
devices we outlined the schemes for a spin filter, spin guide, and a
spin current switch (spin transistor). This program promises to
build spintronics devices avoiding magnetic materials that are not
conventional for the semiconductor industry. It can be realized in
the gated heterostructures with a sufficiently strong Bychkov-Rashba
SO interaction \cite{Nitta1997, Engels1997, Sato2001} by
manipulating the gates. \cite{KSF2003}

The effect of the separation of the trajectories of electrons with different
spin polarizations (chiralities) influenced by the space-varying SO
interaction has the same grounds as the double refraction (birefringence) of
light in uniaxial crystals exploited in optical devices for the polarization
of light. The separation of the trajectories of carriers of different spin
has been observed recently\cite{Rokhinson2004} in the case of a homogeneous
SO interaction as a result of action of a perpendicular magnetic field. The
separation of the trajectories after reflection at a lateral potential
barrier in the presence of the SO interaction has been observed in Ref.~%
\onlinecite{Ohio2004}.

Two facts that can be useful for the purposes of the spintronics have been
found in Ref.~\onlinecite{KSF2003} in the analysis of the spin-dependent
scattering of electrons incident on a lateral interface with a SO
interaction varying in the direction normal to the interface. First, there
exists an interval of outgoing angles $\theta ^{c}<\theta <\pi /2$ where
only electrons with a definite spin chirality can penetrate (see Fig.~\ref%
{fig:refraction}). If it is possible to collect electrons from this
interval, one will have an ideal spin filter. Second, electrons of
this chirality exhibit a total internal reflection for an angle of
incidence in an interval $\varphi ^{c}<\varphi <\pi /2,$ where
$\varphi ^{c}$ is a critical angle of the total internal reflection.
It is clear from these observations that the electrons propagating
at small angles to the lateral interface are most sensitive to the
variation of the magnitude of the spin-orbit interaction, and
therefore such electrons are most appropriate for spin control and
manipulation. Hence one has to find a way to create (and collect)
flows of electrons of high intensity that are almost tangential to
the interface. In this paper we suggest using the diffuse
emission\cite{Milne, Nieuwenhuizen} as a possible solution of this
task.

We show that making one of the two regions connected by the lateral
interface to be diffusive is an effective way to achieve a flat angular
distribution of particles emitted from the diffusive region into the clean
one. The effect of flattening of the angular distribution of the emitted
electrons is the robust property of the diffuse emission which holds despite
the presence of a spin-dependent reflection at the interface. Due to the
flatness of the angular distribution a substantial portion of electrons
propagates at small angles to the interface and these electrons are suitable
for the spin filtration and subsequent spin manipulation. In Fig.~\ref%
{fig:diffuse-filter} a scheme of a device is presented which operates as a
spin filter with a high level of spin-polarization of a filtered current.

The paper is organized as follows. In Section \ref{sec:spinorbit} we present
the results\cite{KSF2003} of the analysis of scattering of electrons at the
lateral interface between the two regions with different magnitudes of the
Bychkov-Rashba term. In Section \ref{sec:kinetics} we analyze the transport
of the 2D electrons in the presence of the Coulomb interaction near the
interface between the diffusive and ballistic regions. In Section \ref%
{sec:milne} we reconsider the problem of the diffuse emission (Milne
problem\cite{Milne, Nieuwenhuizen}) for an arbitrary dimension. In Section %
\ref{sec:spindependentemission} we generalize the technique developed for
the problem of the diffuse emission to the case of the spin-dependent
scattering at the lateral interface. We use a semiclassical approach with
electrons moving along the classically allowed trajectories when the spin of
electrons is the only element treated quantum-mechanically. For this
purpose, we analyze the transport near the interface in terms of the spin
density matrix. In the Summary a scheme of a spin filter device that creates
two beams of electrons of a very high level of spin polarization is
presented.

\section{spin-dependent scattering at the lateral interface}

\label{sec:spinorbit} Consider a two-dimensional electron gas confined in
the $xz$ plane by a potential well in the semiconductor heterostructure.
Generally, the potential well has the shape of an asymmetric triangle, and,
consequently, there is a direction of asymmetry, $\mathbf{\hat{l}}$ ,
perpendicular to the electron gas plane. This leads to the appearance of the
specific spin-orbit interaction term\cite{BychkovRashba84} in the
Hamiltonian, $\alpha (\mathbf{p\times \hat{l}})\mathbf{\sigma}$. We will
consider the case when the parameter $\alpha $ varies along the $x$
direction. The direction of $\mathbf{\hat{l}}$ is chosen as $\mathbf{\hat{l}}%
=-\mathbf{\hat{y}}.$ Generally, the Hamiltonian has the form:
\begin{eqnarray}
H_{R} &=&\frac{1}{2m}p_{x}^{2}+\frac{1}{2m}p_{z}^{2}+B(x)  \notag \\
&+&\frac{1}{2}(\mathbf{\hat{l}\times \sigma})[\alpha (x)\mathbf{p}+\mathbf{p}%
\alpha (x)].  \label{eq:Hamiltonian}
\end{eqnarray}
Here $B(x)$ describes the varying bottom of the conduction band which may be
controlled by gates. The current operator corresponding to this Hamiltonian
contains a spin term: $\mathbf{J}=\frac{\mathbf{p}}{m}+\alpha (x)(\mathbf{%
\hat{l}\times \sigma}).$ The presence of spin in the current operator
implies that in the process of scattering at the lateral interface with
varying $\alpha $ the continuity conditions for the wave function involves
the spin degrees of freedom of the electrons. This makes the electron
scattering at the interface to be spin-dependent.

To diagonalize the Hamiltonian with the Bychkov-Rashba term in the regions
of constant $\alpha $ one has to choose the axis of the spin quantization
along the direction $(\mathbf{\hat{l}\times p})$. Then for the two
chiralities (referred to below as $``+"$ and $``-"$ modes) the dispersion
relations are given by
\begin{equation}
E^{\pm}=\frac{p^{2}}{2m}\pm \alpha p+B.  \label{eq:dispersion}
\end{equation}%
Correspondingly, the momenta of the waves of a given energy $E$ involved in
the scattering at the interface between the two regions with different $%
\alpha$ are determined as follows:
\begin{eqnarray}
p_{SO}^{\pm} &=&m(\sqrt{2(E-B)/m+\alpha ^{2}}\mp \alpha )  \notag \\
&=&mv_{F}(\sqrt{1+\widetilde{\alpha}^{2}}\mp \widetilde{\alpha}).
\label{eq:ppm}
\end{eqnarray}%
Here $v_{F}=\sqrt{2(E-B)/m},$ and we introduce a small dimensionless
parameter $\widetilde{\alpha}=\alpha /v_{F}$ which we will use throughout
the paper. Notice that at a given energy the velocity of electrons of both
chiralities is the same: $v={\partial}E^{\pm}/{\partial}p=v_{F}\sqrt{1+%
\widetilde{\alpha}^{2}}$.

\begin{figure}[h]
\centerline{
    \includegraphics[width=0.5\textwidth]{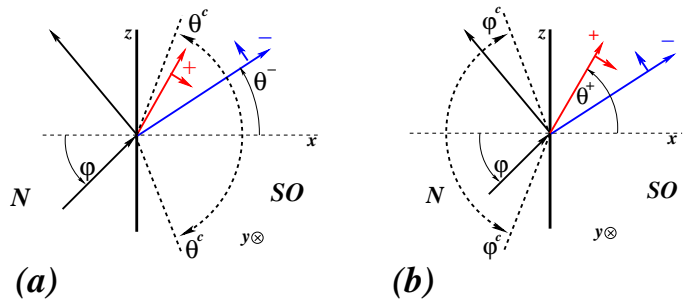}}
\caption{(Color online) The refraction of electrons at the interface between
the regions with (SO) and without (N) spin-orbit interaction. A beam
incident at angle $\protect\varphi $ splits after the refraction into two
beams with $``+"$ and $``-^{\prime\prime}$ chiralities that propagate at
angles $\protect\theta ^{\pm}$ (denoted by red and blue colors,
respectively). (a) $\protect\theta ^{c}$ determines the limited aperture for
$``-"$ chirality; in the angle interval $\protect\theta ^{c}<\protect\theta<%
\protect\pi/2 $ only electrons of $+$ chirality can penetrate. (b) $\protect%
\varphi ^{c}$ is an angle for total internal reflection for electrons of the
$+$ chirality.}
\label{fig:refraction}
\end{figure}

Let us discuss the kinematical aspects of the electron scattering at the
lateral interface. An incident (nonpolarized) beam comes at angle $\varphi$
from the region without a spin-orbit term (or with a suppressed spin-orbit
term), denoted as the N region, and when transmitted into the SO region
splits into two beams of different chirality that propagate at different
angles $\theta^{\pm}$. Figure \ref{fig:refraction} illustrates the
scattering for the simplest case when $\alpha (x<0)=0$. The conservation of
the projection of the momentum on the interface together with Eq.~(\ref%
{eq:ppm}) determine the angles of the transmitted and reflected beams
(Snell's law):

\begin{equation}
p_{N}\sin \varphi =p_{SO}^{\pm}\sin \theta ^{\pm},  \label{eq:Snell}
\end{equation}
where $p_{N}$ is a momentum of an electron in the $N$ region and $%
p_{SO}^{\pm}$ are the momenta in the SO region after passing through the
interface.
From Eqs.~(\ref{eq:ppm}) and (\ref{eq:Snell}) it follows that the SO region
is more ``optically'' dense for the $+$ mode (i.e., it has a smaller wave
vector) and less dense for the $-$ mode. Correspondingly, the $+$ mode is
refracted to larger angles than the $-$ one, and therefore the $+$ mode
exhibits a total internal reflection with a critical angle $\varphi^{c} $.
As to the $-$ mode it has a limited aperture in the SO region for outgoing
angles: $\theta <\theta ^{c}<\pi /2.$

Remarkably, the angle interval where only the $+$ mode can penetrate is not
narrow as its width has a square root dependence on $\widetilde{\alpha}$. It
follows from Snell's law that $(\pi /2-\theta ^{c})\approx (\pi /2-\varphi
^{c})\approx \sqrt{2\widetilde{\alpha}}$. Actually, one can reduce $\theta
^{c}$ even further. With the gates acting selectively on the different
regions of the electron gas, $\delta B=B(-\infty )-B(+\infty )\neq 0$, one
can alter the position of the bands relative to the Fermi level in the N and
SO regions. A simple analysis shows that with an increase of $\delta B$
(i.e., lowering $p_{F}$ in the normal region) the angle interval $(\pi
/2-\theta _{c})$ grows and reaches $2\sqrt{\widetilde{\alpha}}$. Starting
from this moment the angle interval suitable for spin filtration narrows and
eventually becomes $\sim \widetilde{\alpha}$, instead of $\sim \sqrt{%
\widetilde{\alpha}}.$

The problem of scattering of electrons at a lateral interface between the
two regions with different magnitudes of the Bychkov-Rashba term has been
considered in Ref.~\onlinecite{KSF2003} for the two limiting cases of a
sharp, $\lambda /d\gtrsim 1,$ and smooth, $\lambda /d\ll 1,$ interface,
where $\lambda $ is an electron wavelength and $d$ is an effective width of
the interface. Here a qualitative description of the analysis of the
spin-dependent scattering at the interface will be presented only.

A scattering state of an electron coming from the N region in the state $
e^{i(p_{x}x+p_{z}z)}\chi _{\scriptstyle\mathrm{N}}^{+}$ is given by
\begin{widetext}
\begin{equation}
\Psi ^{+}=e^{ip_{z}z}\left\{
\begin{array}{lllll}
e^{ip_{x}x}\chi _{\scriptstyle\mathrm{N}}^{+}+ & e^{-ip_{x}x}\chi _{%
\scriptstyle\mathrm{N}}^{+}r_{++} & + & e^{-ip_{x}x}\chi _{\scriptstyle
\mathrm{N}}^{-}r_{-+}, & \;x<0 \\
& e^{ip_{x}^{+}x}\chi _{\scriptstyle\mathrm{SO}}^{+}t_{++} & + &
e^{ip_{x}^{-}x}\chi _{\scriptstyle\mathrm{SO}}^{-}t_{-+}, & \;x>0%
\end{array}
\right.  \label{eq:Scattering_State}
\end{equation}
\end{widetext}
where $\chi _{\scriptstyle\mathrm{N/SO}}^{\pm}$ are spinors
corresponding to the $\pm $ chiral modes in the $N$ and $SO$
regions, and $r$ and $t$ are the
amplitudes of the reflected and the transmitted waves. In Eq.~(\ref%
{eq:Scattering_State}) the interface is at $x=0$ and for simplicity we limit
ourselves to the case of the interface between the N and SO regions. A
similar expression holds also for $\Psi ^{-}$ which evolves from the
incident state $\chi _{\scriptstyle\mathrm{N}}^{-}.$

The flux of particles impinging on the interface at a given angle $\varphi $
can be defined as $I_{\epsilon}d\epsilon d\varphi ds$, where $ds$ is the
cross-sectional area (width) of a beam. The intensity $I_{\epsilon}^{SO}$of
the transmitted flux of a given chirality can be found from the relation $
\left| \mathcal{T}_{SO\leftarrow N}(\varphi )\right|
^{2}I_{\epsilon}^{N}\cos \varphi d\varphi =I_{\epsilon}^{SO}\cos \theta
d\theta .$ Here the cosine factors take into consideration the change of the
cross-sectional width of the beam as a result of the scattering, while $%
\left| \mathcal{T} _{SO\leftarrow N}(\varphi )\right| ^{2}=\left| t\right|
^{2}(v_{x}^{SO}/v_{x}^{N})$ (we use the fact that for the two chiralities
the velocities are the same). Finally, the intensity of the refracted flux
relative to the incident one is equal to

\begin{equation}
I_{\epsilon}^{SO}/I_{\epsilon}^{N}=(d\theta /d\varphi
)^{-1}(v^{SO}/v^{N})\left| t\right| ^{2}.  \label{eq:intensity}
\end{equation}

For a sharp interface the amplitudes $r,t$ can be found from the continuity
conditions, while for a smooth interface one can conduct the analysis of the
refraction using a small parameter $\eta =(d\alpha /dx)/\alpha p_{F}\sim
\lambda /d\ll 1$. In the latter case the electron spin adjusts itself
adiabatically to the momentum keeping its polarization in the direction
perpendicular to the momentum.

\begin{figure}[h]
\centerline{
  \includegraphics[width=0.3\textwidth]{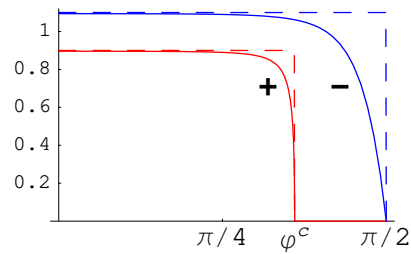}}
\caption{(Color online) The intensities per unit outgoing angle of the
electrons transmitted without change of their chirality $\sim(d \protect%
\theta^+/d \protect\varphi)^{-1} |t_{++}\mid ^{2}$ and $\sim(d \protect\theta%
^-/d \protect\varphi)^{-1} |t_{--}\mid ^{2}$ per unit angle as a function of
an angle of incidence for sharp (solid line) and smooth (dashed line)
interfaces (Ref. \onlinecite{valuealpha}).}
\label{fig:intensity}
\end{figure}

It has been shown in Ref.~\onlinecite{KSF2003} that in the course of
refraction at the interface with $\widetilde{\alpha}\ll 1$ transitions
between states of different chirality are strongly suppressed. For that
reason, the drop of the intensities of the transmitted electrons without
change of their chirality $t_{++}$ and $t_{--}$, see Fig.~\ref{fig:intensity}
, occurs practically only due to the reflection which becomes decisive only
for $\varphi \gtrsim \varphi ^{c}$. In particular, for a smooth interface
the probability of the reflection outside the region of the total internal
reflection is almost entirely suppressed because the matrix elements
describing reflection are integrals of the rapidly oscillating functions.
Consequently, the transmission amplitudes presented in Fig.~\ref%
{fig:intensity} have almost rectangular shape for a smooth interface.

Summarizing the results obtained in Ref.~\onlinecite{KSF2003} for the cases
of the sharp and smooth interface one can state that for both the discussed
cases an electron in a state of a definite chirality propagates along the
classically allowed trajectory for this chirality, while a change of the
chirality is very ineffective.

The case of the N-SO interface described so far was taken mostly for
illustration. Actually, any interface with a change of $\alpha $ results in
splitting of the trajectories that can be used for the purposes of spin
polarization and filtration. In the analysis above, one should replace $
\alpha $ by the difference of the strength of the spin-orbit interaction $
\delta \alpha _{\scriptstyle\mathrm{SO}}$ across the interface.

\section{Kinetic equation, electroneutrality, and boundary conditions}

\label{sec:kinetics} In this section we discuss the kinetics of the
two-dimensional (2D) electron gas near the boundary between the diffusive
and ballistic regions. We show that within the linear response approximation
the equation determining the current distribution function and the equation
determining the density and potential profiles are decoupled, except for
boundary conditions. This analysis justifies the concept introduced by
Landauer:\cite{LandauerIBM} to determine the transport properties it is
enough to consider noninteracting quasiparticles that propagate without
interaction, i.e., ignoring any effects related to the redistribution of the
potential due to the Coulomb interaction of electrons. The problem requires,
however, fixing up the boundary conditions, which we perform below for a
particular geometry.

The kinetic equation describing a stationary flow of electrons by a
distribution function $n_{p}(x)$ is
\begin{equation}
\frac{\partial \epsilon _{p}(\mathbf{r})}{\partial \mathbf{p}}\frac{\partial
n_{p}(\mathbf{r})}{\partial \mathbf{r}}-\frac{\partial \epsilon _{p}(\mathbf{%
r})}{\partial \mathbf{r}}\frac{\partial n_{p}(\mathbf{r})}{\partial \mathbf{p%
}}=\text{St}\{n_{p}(\mathbf{r})\}.  \label{eq:kinetic:1}
\end{equation}
The electron flow is forced by the electric field, $\partial \epsilon _{p}(
\mathbf{r})/\partial \mathbf{r}=e\mathbf{\nabla}\Phi (\mathbf{r})$.
Correspondingly, Eq.~(\ref{eq:kinetic:1}) should be supplied by the Poisson
equation which when limited to the 2D plane is
\begin{equation}
\bigtriangleup \Phi =-4\pi e\rho _{2D}\delta (y).  \label{eq:Poisson}
\end{equation}
Here the Laplacian acts in 3D space, while $\rho _{2D}$ is the deviation of
density of the 2D electron gas from the equilibrium value. Therefore Eq.~(%
\ref{eq:Poisson}) should be supplemented with the 3D conditions on $\Phi $
that will take care of the 3D environment of the 2D electron gas. This
introduces an element of nonuniversality into this problem. Fortunately, in
the linear response approximation the situation is much more tractable and
the problem of the current flow of the 2D gas becomes self-contained (see,
e.g.,~Ref.~\onlinecite{Levinson1989}).

In the linear response approximation one keeps terms linear in $\mathbf{%
\nabla}\Phi (\mathbf{r})$ and $\delta n_{p}(\mathbf{r})$ only:
\begin{equation}
\mathbf{v}_{F}\frac{\partial \delta \overline{n}_{p}(\mathbf{r})}{\partial
\mathbf{r}}-e\mathbf{\nabla}\Phi (\mathbf{r})\mathbf{v}_{F}\frac{\partial
n_{F}^{0}}{\partial \epsilon}=\text{St}\{\delta n\},  \label{eq:linearized}
\end{equation}
where $\delta n_{p}(\mathbf{r})=n_{p}(\mathbf{r})-n_{F}^{0}$ and
$n_{F}^{0}$ is the Fermi-Dirac equilibrium distribution. (We assume
throughout the paper that the spatial variation of the Fermi-energy
level as well as other parameters of the electron liquid such as the
density of states and the screening length occurs on distances
greatly exceeding the wavelength and the mean free path. For that
reason, the spatial variation of the parameters characterizing the
electron liquid is ignored below.) In the presence of the
electron-electron interaction there is an additional force that
originates from the interaction of the quasiparticles. This effect
has been accounted for by a substitution of the local equilibrium
distribution $\delta \overline{n}_{p}(\mathbf{r})$ in place of
$\delta n_{p}(\mathbf{r})$ in Eq.~(\ref{eq:linearized}), see
Ref.~\onlinecite{NozPines}.

Following Refs.~\onlinecite{Wexler} and \onlinecite{Levinson1989}, one can
shift the distribution function by a local value of the potential $\Phi (%
\mathbf{r})$ introducing a displacement function $f(\mathbf{r},\varphi )$
\begin{equation}
\delta \overline{n}_{p}(\mathbf{r})=\frac{\partial n_{F}^{0}}{\partial
\epsilon}e[\Phi (\mathbf{r})-f(\mathbf{r},\varphi )],
\label{eq:displacement}
\end{equation}
where the direction of the momentum $\mathbf{p}$ is given by the angle $
\varphi $. In terms of the displacement function $f(\mathbf{r},\varphi )$
the system of Eqs. (\ref{eq:linearized}) and (\ref{eq:Poisson}) acquires the
form
\begin{eqnarray}
\mathbf{v}_{F}\frac{\partial f(\mathbf{r},\varphi )}{\partial \mathbf{r}}
&=& \text{St}\{f\},  \label{eq:2D-f} \\
\bigtriangleup \Phi -2{\kappa _{2D}}\Phi (\mathbf{r})\delta (y) &=&-2{\kappa
_{2D}}\left\langle f(\mathbf{r},\varphi )\right\rangle _{\varphi}\delta (y),
\label{eq:2D-phi}
\end{eqnarray}
where ${\kappa _{2D}}=2\pi e^{2}\partial n/\partial \mu $ is the inverse
screening length of the 2D electron gas, which in the Thomas-Fermi
approximation is equal to $2e^{2}m$; $\left\langle f(\mathbf{r},\varphi
)\right\rangle _{\varphi}=\int (d\varphi/2\pi)f(\mathbf{r},\varphi ) $ is $f(%
\mathbf{r},\varphi )$ averaged over directions of momentum. Thus, although
Eq.~(\ref{eq:linearized}) together with the Poisson equation (\ref%
{eq:Poisson}) constitutes a system of two coupled equations, the potential $
\Phi (\mathbf{r})$ drops out from Eq.~(\ref{eq:2D-f}) governing the function
$f(\mathbf{r},\varphi )$. Given boundary conditions, the displacement
function $f(\mathbf{r},\varphi )$ determines the current distribution
without any feedback from Eq.~(\ref{eq:2D-phi}):
\begin{equation}
\mathbf{j}(\mathbf{r)}=2e\int \frac{d^{2}p}{(2\pi )^{2}}\mathbf{v}_{F}\delta
\overline{n}_{p}(\mathbf{r})=2e^{2}\nu _{2D}\left\langle \mathbf{v}
_{F}(\varphi )f(\mathbf{r},\varphi )\right\rangle _{\varphi},
\label{eq:currentN-P}
\end{equation}
where $\nu _{2D}$ is the density of states per one spin species, $\nu
_{2D}=m^{\ast}/2\pi .$

When one is interested in the connection of the current density with the
distribution of the potential and density in the 2D electron gas, Eq.~(\ref%
{eq:2D-phi}) should to be involved along with the relation connecting $\rho
_{2D}$ with $\Phi (\mathbf{r})$ and $f(\mathbf{r},\varphi)$:
\begin{eqnarray}
\delta \rho _{2D}(\mathbf{r)} &=&2\int \frac{d^{2}p}{(2\pi )^{2}}\delta {n}
_{p}(\mathbf{r})=  \notag \\
&=&({\kappa _{2D}}/2\pi e)[\left\langle f(\mathbf{r},\varphi )\right\rangle
_{\varphi}-\Phi (\mathbf{r})].  \label{eq:density}
\end{eqnarray}
For good enough conductors the Poisson equation reduces to the condition of
the electroneutrality\cite{kineticsPit} (which is valid in any dimension).
Under these circumstances typical distances on which the potential $\Phi $
in the left-hand side of Eq.~(\ref{eq:2D-phi}) changes are much longer than
the screening length ${\kappa _{2D}^{-1}.}$ Correspondingly, $\big|\nabla
\Phi \big|\ll \big|{\kappa _{2D}}\big\langle f(\mathbf{r},\varphi)\big\rangle%
_{\varphi}\big|,\;\;\big|{\kappa _{2D}}\Phi (\mathbf{r})\big|,$ and the
Poisson equation reduces to the condition of the electroneutrality:
\begin{equation}
\delta \rho _{2D}(\mathbf{r)}=0;\qquad \left\langle f(\mathbf{r},\varphi
)\right\rangle _{\varphi}=\Phi (\mathbf{r}).  \label{eq:neutrality}
\end{equation}

\begin{figure}[t]
\centerline{
\includegraphics[width=0.3\textwidth]{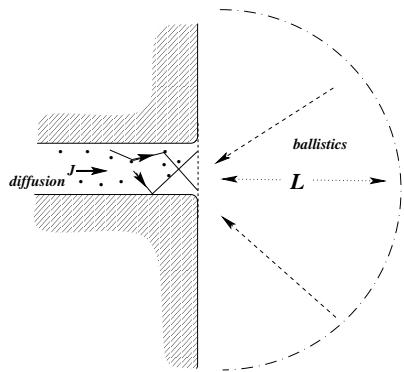}}
\caption{ The physical setup which leads to Eq. (\ref{eq:bcinfinity}) as a
boundary condition for the kinetic equation. The diffusive region has a
geometry of a stripe that runs into a ballistic basin. The current density
vanishes inside the ballistic region at a typical distance $L\ll l_{in}$.
Trajectories that start from a terminal on the ballistic side of the device
and reach the diffusive stripe are shown by dashed lines.}
\label{fig:basin}
\end{figure}

Now we turn to the discussion of the boundary conditions. Let us assume that
the diffusive region has a stripe geometry with the $x$ axis directed along
the stripe (see Fig. \ref{fig:basin}). In the diffusive region the collision
term is controlled by the elastic relaxation time $\tau _{el}$, and the
kinetic equation acquires the form
\begin{eqnarray}
\mathbf{v}_{F}\frac{\partial f(\mathbf{r},\varphi )}{\partial \mathbf{r}}
&=& \text{St}\{f\}_{elastic}  \notag \\
&=&-\frac{f(\mathbf{r},\varphi )-\langle f(\mathbf{r},\varphi )\rangle
_{\varphi}}{\tau _{el}}\;.  \label{eq:kineticelastic}
\end{eqnarray}
Here we assume that the impurity scattering is of short range nature. (The
angular dependence of scattering of electrons by the impurities does not
influence the distribution function of electrons deep inside the diffusive
region, but is important for the angular profile of the diffuse emission.)
With the elastic mean free path $l=v_{F}\tau _{el}$ used as a unit of
length, the kinetic equation can be rewritten in terms of the dimensionless
variables $\zeta =x/l$:

\begin{equation}
-\cos \varphi \frac{\partial}{\partial \zeta}f(\zeta ,\varphi )+f(\zeta
,\varphi )=\langle f(\mathbf{r},\varphi )\rangle _{\varphi}.
\label{eq:homogeneous}
\end{equation}
Here the minus sign in the first term containing $\partial/\partial \zeta $
appears because $\varphi $ is chosen as an angle formed by a momentum with
the direction $-\widehat{x}$ (note that $-\widehat{x}$ directs inwards the
diffusive region).

In the current carrying state electrons deep inside the diffusive region are
distributed according to the Drude form. Correspondingly, at a distance
(counted from the interface) exceeding few $l$ Eq.~(\ref{eq:homogeneous})
has a solution:
\begin{equation}
f_{\scriptstyle\mathrm{Dr}}(\zeta ,\varphi )=-J\pi ^{-1}(\cos \varphi
+\zeta).  \label{eq:Drude}
\end{equation}
Here the factors are fixed in such a way that the current $J(\zeta)$ defined
as
\begin{equation}
J(\zeta )=-2\pi \left\langle \cos \varphi \,f(\zeta ,\varphi )\right\rangle
_{\varphi}  \label{eq:Jcurrent}
\end{equation}
is equal to $J$ for the distribution function $f_{\scriptstyle\mathrm{Dr}%
}(\zeta ,\varphi )$. With the use of Eq.~(\ref{eq:currentN-P}) the physical
current $j\equiv \mathbf{j}_{x}=-2e^{2}\nu _{2D}v_{F}\left\langle \cos
\varphi \,f(\zeta ,\varphi )\right\rangle _{\varphi}$ can be related to the
current $J$ defined in Eq.~(\ref{eq:Jcurrent}) as follows:
\begin{equation}
j=p_{F}(e^{2}/2\pi ^{2})J.  \label{eq:jJrelation}
\end{equation}

To obtain the relation between the current $j$ and the electric field let us
apply a gradient to the both sides of Eq.~(\ref{eq:density}) with the
distribution function $f_{\scriptstyle\mathrm{Dr}}$ used for $f(\mathbf{r}%
,\varphi )$. Then, together with the fact that the electric field $E=-\nabla
\Phi $, one gets the relation:

\begin{equation}
j=\sigma E-D_{ch}\nabla (e\delta \!\rho _{2D}),  \label{eq:Einstein}
\end{equation}
where the conductivity $\sigma =2e^{2}\nu _{2D}(l^{2}/2\tau )$ while the
diffusion coefficient of charge density, $D_{ch},$ corresponds to the
Einstein relation, $D_{ch}=$ $(\partial \mu /\partial n)\sigma /e^{2}$.
Under the condition of the electroneutrality $\delta \rho _{2D}(\mathbf{r}
)=0 $ and $j=\sigma E$. [Actually Eq.~(\ref{eq:Einstein}) does not require
the Drude form, $f_{\scriptstyle\mathrm{Dr}},$ of the distribution. It holds
in the diffusion approximation when the distribution function has only two
first harmonics $f(x,\varphi )=f^{(0)}(x)-f^{(1)}(x)\cos \varphi $ and the
functions $f^{(0)}(x)$ and $f^{(1)}(x)$ vary on scales much exceeding the
mean free path $l.$]

To proceed further with Eq.~(\ref{eq:2D-f}), one has to specify the
distribution of particles incident from the terminals located on the
ballistic side of the device under discussion. Generally, this distribution
is not universal as it depends on a particular geometry. For definiteness,
we consider the case when the diffusive region that has stripe geometry runs
into a ballistic basin, see Fig. \ref{fig:basin} (a stripe is wide enough
allowing the transverse quantization of electrons to be ignored). Following
Ref.~\onlinecite{Levinson1989}, we use the method of characteristics to
determine a distribution of particles that impinge onto the diffusive stripe
from the ballistic region (corresponding family of angles $\varphi $ will be
denoted as $\varphi _{b\rightarrow d}$).

In the ballistic region the collision term is controlled by the inelastic
relaxation time $\tau _{in}$, and the kinetic equation acquires the form
\begin{eqnarray}
\mathbf{v}_{F}\frac{\partial f(\mathbf{r},\varphi )}{\partial \mathbf{r}}
&=& \text{St}\{f\}_{inelastic}  \notag \\
&=&-\frac{f(\mathbf{r},\varphi )-\Phi (\mathbf{r)}}{\tau _{in}}.
\label{eq:kinetballist}
\end{eqnarray}
The solution of Eq.~(\ref{eq:kinetballist}) for the directions $%
\varphi_{b\rightarrow d}$ is given by
\begin{equation}
f(\mathbf{r},\varphi )=\int_{-\infty}^{0}dt\frac{e^{t/\tau _{in}}}{\tau _{in}%
}\Phi[\widetilde{\mathbf{r}}(t)_{\mathbf{r},\varphi}],
\label{eq:characteristics}
\end{equation}
where $\widetilde{\mathbf{r}}(t)_{\mathbf{r},\varphi}$ is the trajectory
(the characteristics) of an electron that starts at the remote past from a
terminal on the ballistic side of the device and reaches the point $\mathbf{r%
}$ with the momentum directed along $\varphi $ at the moment $t=0;$ see
dashed lines in Fig. \ref{fig:basin}. We assume that the current carrying
area widens sharply on the ballistic side of the setup. Correspondingly, the
current density vanishes inside the ballistic region at a typical distance $
L $. Under the condition $L\ll l_{in}=v_{F}\tau _{in}$, the integral in Eq.~(%
\ref{eq:characteristics}) is accumulated at distances where electrons are at
the equilibrium and the potential $\Phi (\mathbf{r})$ is equal to its
equilibrium value at a terminal deep inside the ballistic region, $\Phi
(+\infty ).$ Then, it follows from Eq.~(\ref{eq:characteristics}) that for
incoming directions $f(\mathbf{r},\varphi _{b\rightarrow d})=\Phi (+\infty)$%
.
This provides us with the boundary condition to be imposed at the interface,
$x=0,$ on the incoming part of the function:
\begin{equation}
f(x=0,\varphi _{b\rightarrow d})=\Phi (+\infty ).  \label{eq:bcinfinity}
\end{equation}
The obtained boundary condition is isotropic. This remarkable feature is a
consequence of the choice of the proper geometry.

Together with the current distribution deep in the diffusive region given by
Eq.~(\ref{eq:Drude}), the relation (\ref{eq:bcinfinity}) constitutes the
full set of the boundary conditions needed for the solution of Eq.~(\ref%
{eq:kineticelastic}). Two remarks are now in order to complete the
discussion of the boundary conditions.

(i) Notice that $\Phi (x=0)\neq \Phi (+\infty ).$ The point is that due to
the abrupt change in scattering the solution $f(\mathbf{r},\varphi )$ of
Eq.~(\ref{eq:kineticelastic}) has a singular derivative near the interface.
Under these circumstances, one cannot neglect the term $\bigtriangleup \Phi $
in Eq.~(\ref{eq:2D-phi}), and therefore the condition of the
electroneutrality is violated in the vicinity of the interface in a strip of
a width $\propto {\kappa _{2D}}^{-1}$. The deviation of the density
distribution $\delta \rho _{2D}(\mathbf{r})$ from the equilibrium at the
interface leads to the variation of the potential, and hence $\Phi (x=0)$
differs from $\Phi (+\infty )$.

(ii) As Eq.~(\ref{eq:kineticelastic}) can be satisfied by $f(\mathbf{r}%
,\varphi )=const$, any solution of Eq.~(\ref{eq:kineticelastic}) can be
shifted by a constant with no consequences for the physical quantities [at
any measurement of the current one registers the difference between fluxes
of incoming and outgoing particles and the isotropic part of $f(\mathbf{r}%
,\varphi )$ is cancelled out]. It follows from Eq.~(\ref{eq:displacement})
that the distribution of particles impinging onto the diffusive region at
the interface, $\delta \overline{n}_{\varphi _{b\rightarrow d}}(x=0)$, is
not affected by a global shift of the potential: $\delta \overline{n}%
_{\varphi _{b\rightarrow d}}(x=0)=(\partial n_{F}^{0}/\partial \epsilon)e%
\left[\Phi (x=0)-\Phi (+\infty )\right] $. This is in full accord with the
fact that one is free to shift the potential $\Phi $ by a constant.

In Sec.~\ref{sec:milne} and \ref{sec:spindependentemission} we choose $\Phi
(+\infty )=0$ and correspondingly use $f(x=0,\varphi _{b\rightarrow d})=0$
as the boundary condition for the function $f(\mathbf{r},\varphi )$ at the
interface between the diffusive and ballistic regions.

\section{diffuse emission problem}

\label{sec:milne} 
\begin{figure}[bph]
\centerline{\includegraphics[width=0.3\textwidth]{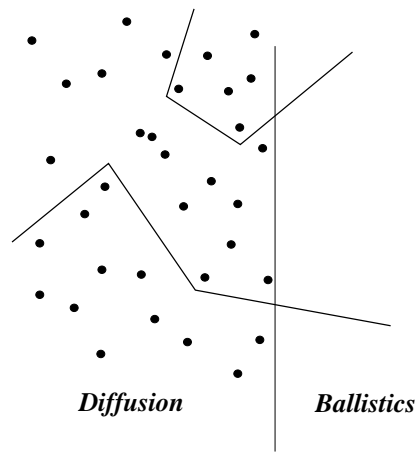}}
\caption{Milne problem. The diffusive region is to the left of the
interface, $\protect\zeta <0$. The ballistic region is to the right, $%
\protect\zeta >0$.}
\label{fig:Milne}
\end{figure}

In this section we present a solution of the
classical problem of the diffuse emission (Milne problem\cite{Milne,
Nieuwenhuizen}) (see Fig.~\ref{fig:Milne}) in the form convenient
for the subsequent analysis of Sec.~\ref{sec:spindependentemission}.
It is given for
an arbitrary dimension $d$, but we are interested in the particular case of $%
d=2$.

Let us specify the notation for angle $\varphi $ used throughout Sec.~\ref%
{sec:milne} and \ref{sec:spindependentemission}: for each of the two
regions, diffusive and ballistic, $\varphi $ is chosen as an angle formed by
the momentum of an electron with a normal to the interface directed inwards
to the corresponding region. Correspondingly, for $\mu =\cos \varphi $ we
adopt the convention\cite{Levinson77} that for electrons propagating away
from the interface $\mu >0$ (denoted as $+\mu $ in what follows), while $\mu
<0$ (denoted $-\mu $) corresponds to electrons propagating towards the
interface.

The general solution of Eq.~(\ref{eq:homogeneous}) can be written as a sum
of the current carrying Drude flow [see Eq.~(\ref{eq:Drude})] and the
currentless ``counterflow'' $g(\zeta ,\mu )$, i.e., $f(\zeta ,\mu )=f_{%
\scriptstyle\mathrm{Dr}}(\zeta ,\mu )+g(\zeta ,\mu )$. At a distance about a
few mean free paths $l$ the counterflow $g(\zeta ,\mu )$ approaches an
isotropic distribution (generally, a nonzero constant). For currentless
counterflow particles injected into the diffusive region eventually return
back to the interface. Hence the distribution function of particles emitted
by the diffusive region, $g(-\mu )$, should be determined completely by the
distribution function of the injected particles, $g(+\mu )$. Due to the
linear nature of Eq.~(\ref{eq:homogeneous}), the $g(\pm\mu )$ parts of the
distribution function are related linearly \cite{Milne} by the angular
Green's function $S(\mu ,\mu ^{\prime})$:
\begin{equation}
g(\zeta ,-\mu )=\frac{1}{\mu}\int_{+}d\Omega ^{\prime}S(\mu ,\mu
^{\prime})g(\zeta ,+\mu ^{\prime}).  \label{eq:counterflow}
\end{equation}
Here $d\Omega ^{\prime}$ stands for the angular integration in $d$
dimensions, while the subscript $+$ in the integral indicates that the
integration is limited to the directions for which $\mu ^{\prime}>0$. With
the use of Eq.~(\ref{eq:counterflow}) one can reexpress the density $\rho
_{g}(\zeta )$ corresponding to $g(\zeta ,\mu )$ through the incoming part of
the distribution only:
\begin{equation}
\rho _{g}(\zeta )=\int_{+}d\Omega \text{}H(\mu )g(\zeta ,+\mu )
\label{eq:density-H}
\end{equation}
where
\begin{equation}
H(\mu )=\left[ 1+\int_{+}d\Omega ^{\prime}\frac{S(\mu ,\mu ^{\prime})}{\mu
^{\prime}}\right] .  \label{eq:functionH}
\end{equation}

The Green's function $S(\mu ,\mu ^{\prime})$ satisfies a nonlinear integral
equation which can be derived as follows. Differentiate Eq.~(\ref%
{eq:counterflow}) with respect to $\zeta $, express the derivatives $
\partial g/\partial \zeta $ through the kinetic equation (\ref%
{eq:homogeneous}), and use Eq.~(\ref{eq:counterflow}) to eliminate $%
g(\zeta,-\mu )$ in favor of $g(\zeta ,+\mu )$. In result one gets
\begin{eqnarray}
\int_{+}d\Omega ^{\prime} &S(\mu ,\mu ^{\prime})&\left( \frac{1}{\mu
^{\prime}}+\frac{1}{\mu}\right) g(\zeta ,+\mu ^{\prime}) \\
&=&\frac{1}{S_{d}}H(\mu )\int_{+}d\Omega ^{\prime}\text{}H(\mu
^{\prime})g(\zeta ,+\mu ^{\prime}),  \notag
\end{eqnarray}
where $S_{d}$ is a total solid angle in $d$-dimensions. As this relation
holds for an arbitrary incident distribution the equation for the Green's
function $S(\mu ,\mu ^{\prime})$ follows:
\begin{equation}
S(\mu ,\mu ^{\prime})\left( \frac{1}{\mu ^{\prime}}+\frac{1}{\mu}\right) =
\frac{1}{S_{d}}H(\mu )H(\mu ^{\prime}).  \label{eq:solution-for-S}
\end{equation}
The counterflow does not carry current and therefore it satisfies
\begin{equation}
\int_{+}d\Omega \text{}\left[ \mu g(+\mu )-\mu g(-\mu )\right] =0.
\label{eq:nocurrent}
\end{equation}
This implies that Green's function $S(\mu ,\mu ^{\prime})$ should satisfy
the condition
\begin{equation}
\mu =\int_{+}d\Omega ^{\prime}S(\mu ,\mu ^{\prime})  \label{eq:condition}
\end{equation}
that can be verified with the use of Eq.~(\ref{eq:solution-for-S}) along
with the relation $\int_{+}d\Omega $ $H(\mu )=S_{d}.$

As it has been explained in Sec.~\ref{sec:kinetics} the geometry of the
discussed setup is such that the distribution function in the ballistic
region does not contain a component describing particles propagating towards
the interface, i.e., $f(0,+\mu )=0.$ Therefore, in the solution $f(\zeta
,\mu )=f_{\scriptstyle\mathrm{Dr}}(\zeta ,\mu )+g_{\scriptstyle\mathrm{Dr}%
}(\zeta ,\mu )$ one has to counterbalance the part $f_{\scriptstyle\mathrm{\
Dr}}(0,+\mu )$ by a proper choice of $g_{\scriptstyle\mathrm{Dr}}(0,+\mu
)=-f_{\scriptstyle\mathrm{Dr}}(0,+\mu )$. Then the emission outside the
diffusive region consists of the two contributions: $f_{\scriptstyle\mathrm{%
Dr}}(0,-\mu )$ from the Drude flow, and $g_{\scriptstyle\mathrm{Dr}}(0,-\mu )
$ which originates from the compensating counterflow. As $g_{\scriptstyle
\mathrm{Dr}}(0,+\mu )$ is known, the latter contribution can be found with
the help of Eq.~(\ref{eq:counterflow}). As a result, the emission in the
Milne problem is given by the expression

\begin{equation}
f_{\scriptstyle\mathrm{M}}(0,-\mu )=J\frac{d}{S_{d}}[\mu +\frac{1}{\mu}
\int_{+}d\Omega ^{\prime}\mu ^{\prime}S(\mu ,\mu ^{\prime})],
\label{eq:pre-milne-emission}
\end{equation}
which can be simplified using Eqs.~(\ref{eq:solution-for-S})~and~(\ref%
{eq:condition}) together with the relation $\int_{+}d\Omega \mu H(\mu
)=S_{d}/\sqrt{d}$. Finally, the angular distribution of the particles
emitted from the interface into the clean region is given by
\begin{equation}
f_{\scriptstyle\mathrm{M}}(0,-\mu )=J\frac{\sqrt{d}}{S_{d}}H(\mu ).
\label{eq:milne-emission}
\end{equation}
This result coincides with the one presented in Ref.~\onlinecite{Milne} for
the case $d=3$.
\begin{figure}[h]
\centerline{
    \includegraphics[width=0.3\textwidth]{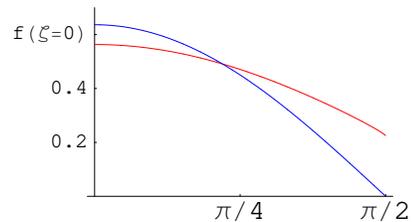}}
\caption{(Color online) The angular distribution, $H(\cos \protect\varphi)$,
of the diffuse emission (red line) as compared to the Drude flow
distribution (blue line) for $J=1$. The function $H(\cos\protect\varphi)\neq
0$ at $\protect\varphi=\protect\pi/2$.}
\label{fig:diffuse_emission}
\end{figure}
In Fig.~\ref{fig:diffuse_emission} the function $H(\mu)$ and the angular
dependence of the Drude flow distribution, $\sim \cos \varphi$, are plotted,
both normalized to $J=1$. As compared to the Drude flow, the diffuse
emission distribution in the Milne problem has a qualitatively different
behavior at large angles. Namely, the distribution flattens and a
considerable part of the distribution is transferred to the large angles.

\section{Diffuse emission in the presence of spin-dependent scattering at
the N/SO interface.}

\label{sec:spindependentemission} In this section we generalize the solution
of the Milne problem to the case of spin-dependent reflection at the
interface. We are mainly concerned with the influence of the strong
reflection at the angles tangential to the interface on the distribution of
the emitted electrons. We show that the effect of flattening of the angular
distribution is the robust property of the solution of the Milne problem
which becomes even stronger in the presence of such a reflection.

\begin{figure}[bph]
\centerline{\includegraphics[width=0.3\textwidth]{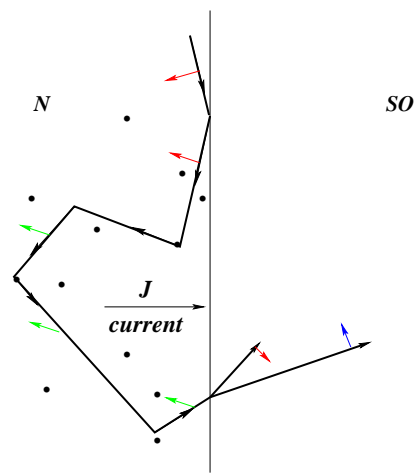}}
\caption{(Color online) Milne problem in the presence of the spin-dependent
scattering at the interface. The SO interaction in the diffusive region ($%
\protect\zeta<0$) is suppressed.}
\label{fig:MilneSO}
\end{figure}

We now concentrate on the calculation of the diffuse emission in the
presence of the spin-orbit interaction at the ballistic side of the
junction. As compared to the consideration in the previous section the
scattering at the interface now depends on the direction of spin of the
incoming electron. On the other hand, electrons after a sequence of random
scattering in the diffusive region return to the interface preserving their
spin (see Fig.~\ref{fig:MilneSO}), as the impurity scattering of electrons
in the diffusive region is assumed to be spin-independent.

To describe such scattering, one should introduce a spin density matrix ${%
\mathbf{\varrho}}^{\varsigma\varsigma ^{\prime}}$. In the diffusive region
each component of the spin density matrix satisfies the kinetic equation (%
\ref{eq:homogeneous}) because in the normal region the Hamiltonian is
spin-independent. Consequently, one can apply the same angular Green's
function $S(\mu ,\mu ^{\prime})$ as in Eq.~(\ref{eq:counterflow}).

We have to describe the counterflow at the interface in the presence of the
reflection. For clarification purposes, let us imagine an auxiliary line, $
\zeta =-0,$ separating the diffusive region from the N-SO interface.
(Actually, for a real device there can be a physical, not auxiliary,
interval between the diffusion region and the beginning of the interface
where $\alpha $ and $B$ start to vary.) Similarly to the discussion of the
Milne problem it is assumed that the electron distribution is given by a sum
of the Drude flow and a counterflow. The incoming part of the counterflow
consists of the contributions: one, $\mathbf{\upsilon}_{\scriptstyle
\mathrm{Dr}}(\zeta =-0,+\mu ),$ counterbalances the incoming part of the
Drude flow, $\mathbf{\upsilon}_{\scriptstyle\mathrm{Dr}}(\zeta
=-0,+\mu)=-\varrho _{\scriptstyle\mathrm{Dr}}(\zeta =-0,+\mu )$, while the
other, $\mathbf{\upsilon}_{refl}(\zeta =-0,+\mu)$, is determined by the flow
incident onto the line $\zeta =-0.$ As it has been explained is Sec.~\ref%
{sec:kinetics} and has been already used in Eq.~(\ref{eq:pre-milne-emission}%
) we assume that there are no electrons incident onto the N-SO interface
from the SO region. In the absence of electrons incident from the SO side of
the interface the flow incident onto the line $\zeta =-0$ comes only as a
result of the reflection at the N-SO interface [the latter circumstance
explains our choice of a subscript for $\mathbf{\upsilon}_{refl}(\zeta
=-0,+\mu )$]. Therefore the overall distribution of particles coming from
the interface $\mathbf{\varrho}_{overall}(\zeta=-0,+\mu)=\mathbf{\upsilon}%
_{refl}(\zeta =-0,+\mu)$.

As a result of scattering inside the diffusive region $\mathbf{\upsilon}_{%
\scriptstyle\mathrm{Dr}}(\zeta =-0,+\mu )$ transforms into a part of the
outgoing counterflow $\mathbf{\upsilon}_{\scriptstyle\mathrm{Dr}}(\zeta
=-0,-\mu ).$ Together with the Drude flow it yields the following
contribution to the outgoing distribution $\varrho _{\scriptstyle\mathrm{M}%
}(\zeta =-0,-\mu )=J(\sqrt{d}/S_{d})H(\mu )\sigma _{0}$, where $\sigma _{0}$
is the $2\times 2$ unit matrix. The overall distribution of particles
emitted from the line $\zeta =-0$ and incident onto the N-SO interface is
given by
\begin{align}
{\mathbf{\varrho}}_{overall}(\zeta =-0,-\mu )=&J\frac{\sqrt{d}}{S_{d}} H(\mu
)\sigma _{0} +\frac{1}{\mu}\int_{+}d\Omega ^{\prime}\,S(\mu ,\mu ^{\prime})
\notag \\
\times&{\mathbf{\upsilon}_{refl}}(\zeta =-0,+\mu ^{\prime}).
\label{eq:emission-spin}
\end{align}
The second term in this equation is generated by the incident part of the
distribution, $\mathbf{\upsilon}{_{refl}}(\zeta =-0,+\mu ),$ which in turn
is determined by the reflection of $\varrho _{overall}(\zeta =-0,-\mu )$ at
the N-SO interface. The relation between the incident and reflected parts of
the distribution should be found from the solution of the scattering problem
at the normal side of interface
\begin{equation}
\mathbf{\upsilon}_{refl}(\zeta =-0,+\mu )={\mathcal{R}}(\mu ){\mathbf{\varrho%
}}_{overall}(\zeta =-0,-\mu ){\mathcal{R}}^{\dagger}(\mu ),
\label{eq:bc-refl-spin}
\end{equation}
where by $\mathcal{R}$ we denote a $2\times 2$ block of the scattering
matrix corresponding to the reflection at the normal side of the interface.

After substituting Eq.~(\ref{eq:bc-refl-spin}) in Eq.~(\ref{eq:emission-spin}%
) one obtains a closed equation for ${\mathbf{\varrho}}_{overall}(\mu)$:
\begin{align}
{\mathbf{\varrho}}_{overall}(-\mu ) =&J\frac{1}{\sqrt{2}\pi}H(\mu)\sigma
_{0}+ \frac{1}{\mu}\int_{+}d\Omega ^{\prime}S(\mu ,\mu ^{\prime})  \notag \\
\times&{\mathcal{R}}(\mu ^{\prime}){\mathbf{\varrho}}_{overall}(-\mu
^{\prime}){\mathcal{R}} ^{\dagger}(\mu ^{\prime}).
\label{eq:emission-spin-with-bc}
\end{align}

To analyze the effect of the reflection at the interface on the form of the
diffuse emission, we study Eq.~(\ref{eq:emission-spin-with-bc}) for the case
of a smooth interface. In this limit the kinematical aspect of the
scattering at the interface is most pronounced and not masked by unnecessary
complications. Namely, we assume that for electrons of the $-$ chirality the
probability of the transmission is $1$ at all angles, while for electrons of
the $+$ chirality the probability of the transmission is $1$ for $\varphi
<\varphi ^{c}$ and $0$ otherwise; see Fig.~\ref{fig:intensity} and the
related discussion of the smooth interface.

We will use a notation $|\,\chi ^{\pm}(\pm \mu ,i)\,\rangle$ for spinors.
The sign of the first argument indicates the sign of the projection of the
momentum of a scattering electron on the direction perpendicular to the
interface, while $i=\pm 1$ is the sign of the momentum component along the
interface; the superscript $\pm $ denotes the chirality. Also in what
follows the integration over $\varphi $ will be performed as $\int d\varphi
=\sum_{i=\pm 1}\int_{-1}^{1}(d\mu/\sqrt{1-\mu ^{2}})$. In this notation the
reflection part of the scattering matrix for the smooth interface is given
by
\begin{equation}
\begin{array}{llcl}
{\mathcal{R}} & = 0, & \;\varphi <\varphi ^{c} &  \\
& =\sum_{i}|\,\chi ^{+}(+\mu ,i)\,\rangle \;\langle \,\chi ^{+}(-\mu ,i)\,|,
& \; \varphi >\varphi ^{c}\; . &
\end{array}
\label{eq:reflection}
\end{equation}

For each direction $-\mu $ we introduce the orthogonal basis $\mathbf{n}
^{\alpha}(-\mu ,i)$ in such a way that $\mathbf{n}^{1}$ is parallel to the
incident momentum $\mathbf{p}$; $\mathbf{n}^{2}$ is perpendicular to the
electron gas plane, $\mathbf{n}^{2}\parallel \mathbf{\hat{l}};$ and $\;
\mathbf{\ n}^{3}$ is directed along the vector of the polarization of the $+$
chirality. For the analysis of Eq.~(\ref{eq:emission-spin-with-bc}) it will
be convenient to use the four-component Bloch vector $(s^{0},\mathbf{s})$
related to the matrix ${\mathbf{\varrho}}$ as follows:
\begin{eqnarray}
{s}^{0}(-\mu ,i) &=&\frac{1}{2}\mathrm{Tr}\sigma ^{0}{\mathbf{\varrho}}(-\mu
,i),  \notag \\
{s}^{\alpha}(-\mu ,i) &=&\frac{1}{2}\mathrm{Tr}(\mathbf{n}^{\alpha}\mathbf{%
\;\sigma}){\mathbf{\varrho}}(-\mu ,i)  \notag \\
{\mathbf{\varrho}}(-\mu ,i) &=&s^{0}\sigma ^{0}+\sum_{\alpha}{s}%
^{\alpha}(-\mu ,i)(\mathbf{n}^{\alpha}\mathbf{\sigma}),  \label{eq:Bloch}
\end{eqnarray}
where $\sigma ^{0}$ is the unit matrix and $\mathbf{\sigma}$ are the Pauli
matrices. In terms of the $(s^{0},\mathbf{s})$ matrix Eq.~(\ref%
{eq:emission-spin-with-bc}) can be rewritten as a set of coupled equations
\begin{align}
{s}^{0}(-\mu ,i) =&J\frac{1}{\sqrt{2}\pi}H(\mu ) +\frac{1}{\mu}%
\sum_{j}\int_{0}^{\cos \varphi ^{c}}\frac{d\mu ^{\prime}}{\sqrt{1-\mu
^{\prime 2}}}S(\mu ,\mu ^{\prime})  \notag \\
\times&\frac{1}{2}\big[\,{s}^{0}(-\mu ^{\prime},j)+{s}^{3}(-\mu
^{\prime},j)\,\big]\;,  \label{eq:emission-spin-components-0}
\end{align}
\begin{align}
{s}^{\alpha}(-\mu ,i)=&\frac{1}{\mu}\sum_{j}\int_{0}^{\cos \varphi ^{c}}
\frac{d\mu ^{\prime}}{\sqrt{1-\mu ^{\prime 2}}}S(\mu ,\mu ^{\prime})  \notag
\\
\times &\langle\,\chi ^{+}(+\mu ^{\prime},j)\,|\frac{1}{2}\mathbf{\ \sigma}%
\cdot \mathbf{n}^{\alpha}(-\mu ,i)|\,\chi ^{+}(+\mu ^{\prime},j)\,\rangle
\notag \\
\times &\big[\,{s}^{0}(-\mu ^{\prime},j)+{s}^{3}(-\mu ^{\prime},j)\,\big].
\label{eq:emission-spin-components-alf}
\end{align}
Remarkably, all components ${s}^{\alpha}$ are determined by a single
combination $f_{\scriptstyle\mathrm{N}}^{+}(-\mu ,j)={s}^{0}(-\mu ,j)+{s}
^{3}(-\mu ,j)$. The combinations ${s}^{0}(-\mu ,j)\pm {s}^{3}(-\mu ,j)$
describe the probability for an incident electron to be in the $\pm $ chiral
state. The representation of $f_{\scriptstyle\mathrm{N}}^{\pm}$ as ${s}
^{0}\pm {s}^{3}$ is a consequence of the choice of the axis $\mathbf{n}^{3}$
along the direction of the spin polarization of an electron in the state of
the $+$ chirality. The fact that the right-hand sides of Eqs.~(\ref%
{eq:emission-spin-components-0})~and~(\ref{eq:emission-spin-components-alf})
depend solely on $f_{\scriptstyle\mathrm{N}}^{+}(-\mu ^{\prime},j)$ has a
simple physical reason: electrons of this chirality only are reflected at
the interface. [In Eq.~(\ref{eq:emission-spin-components-alf}) the component
$s^{1}$ is determined by $f_{\scriptstyle\mathrm{N}}^{+}(-\mu ^{\prime},j)$
without any feedback, and $s^{2}$ vanishes identically in the chosen
parametrization.]

To proceed further we have to calculate $\langle \,\chi ^{+}(+\mu
^{\prime},j)\,|\frac{1}{2}\mathbf{\sigma}\cdot\mathbf{n}^{\alpha}(-\mu
,i)|\,\chi ^{+}(+\mu ^{\prime},j)\,\rangle $ for both $i,j=\pm 1$.
With the use of the relation that determines the direction of the
polarization of a spinor, $ \langle \,\chi ^{+}(+\mu
^{\prime},j)\,|\mathbf{\sigma}|\,\chi ^{+}(+\mu
^{\prime},j)\,\rangle =\mathbf{n}_{{\chi}^{+}(+\mu ^{\prime},j)}$,
the discussed expression is reduced to $\frac{1}{2}\mathbf{n}_{\chi
^{+}(+\mu ^{\prime},j)}\cdot \mathbf{n}^{\alpha}(-\mu ,i)$. Then Eqs.~(\ref%
{eq:emission-spin-components-0})~and~(\ref{eq:emission-spin-components-alf}
), can be rewritten in the form:
\begin{align}
f_{\scriptstyle\mathrm{N}}^{+}(-\mu, i=\pm 1)=&J\frac{1}{\sqrt{2}\pi}H(\mu )+%
\frac{1}{\mu}\int_{0}^{\cos \varphi ^{c}}\frac{d\mu ^{\prime}}{\sqrt{1-{\mu
^{\prime}}^{2}}}S(\mu ,\mu ^{\prime})  \notag \\
\times&\frac{1}{2}\big\{\;\,[1-\cos (\;\;\varphi ^{\prime}\pm \varphi )]f_{%
\scriptstyle\mathrm{N}}^{+}(-\mu ^{\prime},+1)  \notag \\
+&[1-\cos (-\varphi ^{\prime}\pm \varphi )]f_{\scriptstyle\mathrm{N}%
}^{+}(-\mu ^{\prime},-1)\big\},  \label{eq:emission-i}
\end{align}
where angles $\varphi $ and $\varphi ^{\prime}$ are defined in Fig.~\ref%
{fig:angles}. For the difference $\delta f_{\scriptstyle\mathrm{N}}^{+}(-\mu
)=f_{N}^{+}(-\mu ,+1)-f_{\scriptstyle\mathrm{N}}^{+}(-\mu ,-1)$ one gets a
homogeneous equation
\begin{figure}[t]
\centerline{ \includegraphics[width=0.3\textwidth]{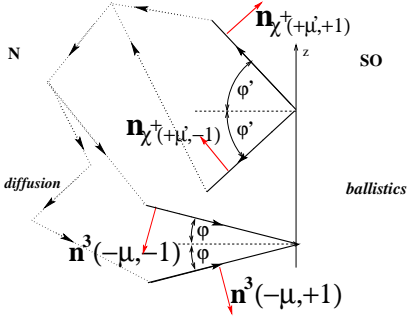}}
\caption{(Color online) Angles and directions of spinors involved in the
derivation of Eq.~(\ref{eq:emission-i}). For each angle $\protect\varphi $
and $ \protect\varphi ^{\prime}$ there are two possible momentum directions
corresponding to a different sign of the $p_{z}$ component.}
\label{fig:angles}
\end{figure}
\begin{eqnarray}
\delta f_{\scriptstyle\mathrm{N}}^{+}(-\mu ) &=&\frac{1}{\mu}\int_{0}^{\cos
\varphi ^{c}}\frac{d\mu ^{\prime}}{\sqrt{1-{\mu ^{\prime}}^{2}}}S(\mu ,\mu
^{\prime})\times  \label{eq:emission-delta-f-N} \\
&\times &\frac{1}{2}[-\cos (\varphi ^{\prime}+\varphi )+\cos (\varphi
^{\prime}-\varphi )]\delta f_{\scriptstyle\mathrm{N}}^{+}(-\mu ^{\prime}).
\notag
\end{eqnarray}
One can check that all eigenvalues of the right-hand side of this equation
considered as a kernel of the transformation are less than 1. Therefore we
conclude that $\delta f_{\scriptstyle\mathrm{N}}^{+}(-\mu )=0$, i.e., $f_{%
\scriptstyle\mathrm{N}}^{+}(-\mu ,i)$ is a symmetrical function with respect
to $i$. Finally, this yields for Eq.~(\ref{eq:emission-i})
\begin{align}
f_{\scriptstyle\mathrm{N}}^{+}(-\mu )=&J\frac{1}{\sqrt{2}\pi}H(\mu ) +\frac{1%
}{\mu}\int_{0}^{\cos \varphi ^{c}}\frac{d\mu ^{\prime}}{\sqrt{ 1-\mu
^{^{\prime}2}}}S(\mu ,\mu ^{\prime})  \notag \\
&\times(1-\mu \mu
^{\prime})f_{\scriptstyle\mathrm{N}}^{+}(-\mu^{\prime}).
\label{eq:numerics}
\end{align}

Following the same route, it can be found from Eq.~(\ref%
{eq:emission-spin-components-alf}) the distribution of particles in the $-$
chiral state:
\begin{align}
f_{\scriptstyle\mathrm{N}}^{-}(-\mu )=& J\frac{1}{\sqrt{2}\pi}H(\mu)+\frac{1%
}{\mu}\int_{0}^{\cos \varphi _{c}}\frac{d\mu ^{\prime}}{\sqrt{ 1-\mu
^{^{\prime}2}}}S(\mu ,\mu ^{\prime})  \notag \\
&\times (1+\mu \mu ^{\prime})f_{\scriptstyle\mathrm{N}}^{+}(-\mu ^{\prime}).
\label{eq:numericsminus}
\end{align}
Equations~(\ref{eq:numerics}) and (\ref{eq:numericsminus}) can be analyzed
using the smallness of $\cos \varphi ^{c}$, but they can be easily solved
numerically. In what follows we will use the results of numerical analysis
for the distribution functions $f_{\scriptstyle\mathrm{N}}^{\pm}(-\mu )$.

\begin{figure}[t]
\centerline{
    \includegraphics[width=0.3\textwidth]{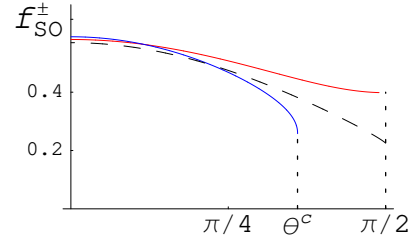}}
\caption{(Color online) The displacement function $f_{\scriptstyle\mathrm{SO}%
}^{\pm}( \protect\theta)$ of the $``+"$ (shown in red) and ``$-$''(shown in
blue) polarization components of the transmitted electrons for $\widetilde{%
\protect\alpha}=0.1$ as compared with the solution without reflection, $%
\widetilde{\protect\alpha} =0$ (dashed line).}
\label{fig:transmission}
\end{figure}

Ultimately, we are interested in the distributions on the SO side of the
interface. To achieve this goal, we have to connect the calculated
distributions $f_{\scriptstyle\mathrm{N}}^{\pm}$ to the distributions on the
SO side $f_{\scriptstyle\mathrm{SO}}^{\pm}$. It follows straightforwardly
from the Liouville's theorem that
\begin{equation}
f_{\scriptstyle\mathrm{SO}}^{\pm}(\epsilon ,+\mu _{\theta})=f_{\scriptstyle
\mathrm{\ N}}^{\pm}(\epsilon ,-\mu _{\varphi}),  \label{eq:Liouville}
\end{equation}
where $\mu _{\theta}$ and $-\mu _{\varphi}$ are two directions connected by
the Snell's law, Eq.~(\ref{eq:Snell}). In the subsequent discussion the
notations $f_{\scriptstyle\mathrm{SO}}^{\pm}(\theta )$ and $f_{\scriptstyle
\mathrm{N}}^{\pm}(\varphi )$ will be used instead of $f_{\scriptstyle
\mathrm{SO}}^{\pm}(\epsilon ,+\mu _{\theta})$ and $f_{\scriptstyle\mathrm{N}%
}^{\pm}(\epsilon ,-\mu _{\varphi});$ notice that we return to the definition
of angles $\varphi $ and $\theta $ as it is given in Fig.~\ref%
{fig:refraction}. Finally, in Fig.~\ref{fig:transmission} our result for $f_{%
\scriptstyle\mathrm{SO}}^{\pm}(\theta )$ is presented.

We observe that the distribution of particles of the $+$ chirality emitted
from the diffusive region into the clean one is flat even in the presence of
the spin-dependent reflection at the interface. Moreover, at large angles
the function $f_{\scriptstyle\mathrm{SO}}^{+}(\theta )$ is noticeably
increased.

As a result of the reflection at the interface there is a redistribution of
the population of the particles of the $+$ and $-$ chiralities.
Consequently, the two chiral components of the current, $j^{\pm},$ change in
the vicinity of the interface, and we are faced with the question of the
conservation of the total current $j=j^{+}+j^{-}$. (We assume for a moment
the width of the diffusive stripe $W$ to be unity, and until the end of this
section will not distinguish between the density of current and the total
current.) To check the current conservation we first calculated numerically
the currents $j^{\pm}$ with the use of the distribution functions $f_{%
\scriptstyle\mathrm{N}}^{\pm}(\epsilon ,-\mu ).$ We get that on the N side
of the interface the current $j^{+}+j^{-}$ is equal to the total current $j$
deep inside the diffusive region. Further on, it can be shown that the two
parts of the physical current $j^{\pm}$ do not change after passing the
interface. To do this notice that in the SO region the chiral components of
the physical current are

\begin{equation}
j_{\scriptstyle\mathrm{\ SO}}^{\pm}=p_{SO}^{\pm}\frac{e^{2}}{2\pi}
\left\langle \cos \theta f_{\scriptstyle\mathrm{\ SO}}^{\pm}(\theta
)\right\rangle  \label{eq:SOcurrent}
\end{equation}
[for the definition of the physical currents see Eqs.~(\ref{eq:Jcurrent})
and (\ref{eq:jJrelation})]. Having in mind Eq.~(\ref{eq:Liouville}) and the
Snell's law in its differential form, $p_{SO}\cos \theta d\theta =p_{N}\cos
\varphi d\varphi$, the following relations for the $\pm $ chirality current
components on the two sides of the interface can be obtained:

\begin{eqnarray}
j^{+} &=&p_{SO}^{+}\frac{e^{2}}{2\pi ^{2}}\int_{0}^{\pi /2}d\theta \cos
\theta f_{\scriptstyle\mathrm{SO}}^{+}(\theta )=  \notag \\
&=&p_{N}\frac{e^{2}}{2\pi ^{2}}\int_{0}^{\varphi ^{c}}d\varphi \cos \varphi
f_{\scriptstyle\mathrm{N}}^{+}(\varphi ),  \label{eq:jplus}
\end{eqnarray}
and

\begin{eqnarray}
j^{-} &=&p_{SO}^{-}\frac{e^{2}}{2\pi ^{2}}\int_{0}^{\theta ^{c}}d\theta \cos
\theta f_{\scriptstyle\mathrm{SO}}^{-}(\theta )=  \notag \\
&=&p_{N}\frac{e^{2}}{2\pi ^{2}}\int_{0}^{\pi /2}d\varphi \cos \varphi f_{%
\scriptstyle\mathrm{N}}^{-}(\varphi ).  \label{eq:jminus}
\end{eqnarray}
For the $+$ chirality component it has to be taken into consideration that
the current of the particles impinging on the interface in the interval of
angles $\varphi ^{c}<\varphi <\pi /2$ is canceled out by the flow of the
reflected particles.

Thus we arrive at the conclusion that the two parts of the physical current $%
j^{\pm}$ do not change after passing the interface, i.e., $j_{\scriptstyle%
\mathrm{N}}^{\pm}=j_{\scriptstyle\mathrm{SO}}^{\pm}$. Together with the fact
of the current conservation in the N region this implies that the total
current in the ballistic SO region is equal to the current deep inside the
diffusive region.

So far we limit ourselves to the diagonal elements of the density matrix.
This was possible because in the N region the density matrix is diagonal in
momentum space, while the current operator is diagonal in spin space. The
situation is different in the SO region where the current operator acquires
the spin structure, see Sec.~\ref{sec:spinorbit}. In addition, the density
matrix becomes nondiagonal in momentum space as an electron beam splits into
two beams as a result of the refraction at the interface with the
inhomogeneous spin-orbit interaction. We will not discuss this problem for
the following reason. As it has been explained in Sec.~\ref{sec:kinetics}
our analysis refers to the case when the current is registered far away from
the near-field zone of the orifice, i.e., at a distance $L$ much exceeding
the width of the diffusive stripe. At such distances the trajectories of
electrons of different chiralities diverge from each other and therefore the
nondiagonal in momentum space components of the density matrix become
nonlocal in space. Since the current operator is local this nonlocality of
the density matrix cannot show up in the current measurements.
\begin{figure}[t]
\centerline{\includegraphics[width=0.3\textwidth]{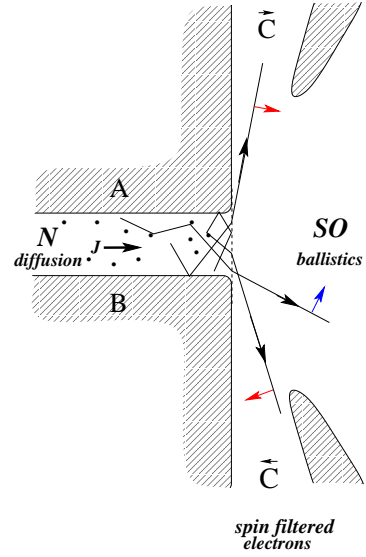}}
\caption{(Color online) A spin filter. Electrons emitted from the diffusive
stripe at small angles to the interface are spin polarized and can be
collected for subsequent spin manipulations.}
\label{fig:diffuse-filter}
\end{figure}

\section{summary}

We have analyzed the transport near the interface with the
inhomogeneous spin-orbit interaction in terms of the spin density
matrix. The present analysis has been performed under the assumption
that the SO interaction in the diffusive region is absent
(suppressed completely). In fact, a much weaker condition is
sufficient. It is enough that spin relaxation rate in the diffusive
region is controlled by the Dyakonov-Perel
mechanism:\cite{DyakonovPerel} $1/\tau_s\sim\Delta_{SO}^2\tau\ll
1/\tau$, where $\Delta_{SO}=2\alpha(x<0)\,p_F\ll1/\tau$ is the spin
splitting of the energy spectrum in the diffusive region, $x<0$.
Under this condition the spin relaxation of the electrons during the
propagation in the diffusive region after the reflection from the
interface is negligible. (The other limit $\Delta_{SO}\agt 1/\tau$
will be considered elsewhere.)

We have verified that the specific property of the solution of the
Milne problem -- the existence of the flat distribution with a large
portion of electrons emitted at small angles to the interface --
still holds in the presence of the reflection at the interface.
Moreover, at large angles the distribution
function of electrons of the $+$ chirality is noticeably increased, see Fig.~%
\ref{fig:transmission}. As it has already been pointed out in Sec.~\ref%
{sec:spinorbit}, there exists an interval of the outgoing angles, $%
\theta^{c}<\theta<\pi/2$, where only the $+$ spin chiral component can
penetrate. Together, these observations call upon to exploit the diffuse
emission for the purposes of spintronics.

In Fig.~\ref{fig:diffuse-filter} a scheme of a device is presented which can
operate as a spin filter for a current passing through the diffusive stripe
confined between nonconducting areas A and B. Two additional barriers (or
gates) are set at a distance $L$ much larger than the width of the diffusive
stripe $W$. Within this geometry each collector, $\overrightarrow{C}$ and $
\overleftarrow{C},$ gets spin carriers emitted into the corresponding
angular interval $\delta \theta $. When $\delta \theta <\pi /2-\theta _{c}$,
particles of the $+$ chirality only can get into the collectors. As a
result, the currents inside each of the two collectors are spin-polarized,
dominantly along the $\pm\,\widehat{x}$ directions.

In the setup under discussion the orifice of the stripe acts as a source of
a current with an angular distribution $f_{\scriptstyle\mathrm{\ SO}%
}^{\pm}(\theta )$. The number of electrons of a certain chirality flowing in
a direction $\theta $ per unit time (i.e. the angular flux) is equal to $
\mathcal{I}_{\scriptstyle\mathrm{\ SO}}^{\pm}(\theta )d\epsilon d\theta $,
where the intensity $\mathcal{I}_{\scriptstyle\mathrm{\ SO}}^{\pm}(\theta )$
is related to $f_{\scriptstyle\mathrm{\ SO}}^{\pm}(\theta )$ as $\mathcal{I}
_{\scriptstyle\mathrm{\ SO}}^{\pm}=\frac{e^{2}}{(2\pi )^{2}}p_{SO}^{\pm}\,f_{%
\scriptstyle\mathrm{\ SO}}^{\pm}(\theta )\,\cos\theta\,W$. The factor $\cos
\theta \,W$ appears because we consider the flux of the particles outgoing
from the orifice at an angle $\theta $. The angular dependence of $\mathcal{I%
}_{\scriptstyle\mathrm{\ SO}}^{\pm}(\theta )$ is in full accord with the
expressions for the density of the currents $j^{\pm} $; see expressions
under the integrals in Eqs.~(\ref{eq:jplus}) and (\ref{eq:jminus}). Assuming
the angular distribution of the emitted electrons to be practically flat,
the spin-polarized current that can be collected by each of the collectors
is $\simeq j\,\delta \theta ^{2}/8$. As it has been explained in Sec.~\ref%
{sec:spinorbit} the width of the angular interval $\delta \theta
=\pi/2-\theta ^{c}$ can be as much as $2\sqrt{\widetilde{\alpha}}$. At this
point the fraction of spin-polarized current reaches its optimal value $%
\simeq j\,\widetilde{\alpha}/2.$ All collected electrons have the same
chirality that results in a very high level of spin polarization of the
current in the collectors. A deviation from the perfect level of
spin-polarization is only due to a small spread of the direction of motion
of electrons within an angular interval $\delta \theta $.

In Ref.~\onlinecite{Sato2001} a large spontaneous spin splitting has been
detected in a gate controlled electron gas formed at a In$_{0.75}$Ga$_{0.25}$
As/In$_{0.75}$Al$_{0.25}$As heterojunction. The reported splitting
corresponds to $\widetilde{\alpha}\approx 0.1.$ For such values of $
\widetilde{\alpha}$ one may expect a rather large angular interval $\delta
\theta $ that can be used for spin filtration; $\delta \theta \approx
36^{\circ}$. Under these conditions, the device proposed in Fig.~\ref%
{fig:diffuse-filter} has the following specifications: a fraction up to $5\%$
of the total current is collected in $\overrightarrow{C}$ and is (almost)
fully spin-polarized along the direction $\widehat{x}$, while the other
fraction of $5\%$ is collected in $\overleftarrow{C}$ and is spin-polarized
along the direction $-\widehat{x}$.

After filtration the spin-polarized current can be manipulated similarly to
the polarized light in optical devices. In particular, one can link the spin
filter to the switch of the spin-polarized current discussed in Ref. %
\onlinecite{KSF2003}.

\begin{acknowledgments}
We thank Y.~B.~Levinson for numerous illuminating discussions.
\end{acknowledgments}


\begin{thebibliography}{99}
\bibitem{KSF2003} M.~Khodas, A.~Shekhter, and A.~M.~Finkel'stein, \emph{%
Phys. Rev. Lett.} \textbf{92}, 086602 (2004).

\bibitem{BychkovRashba84} E.~I.~Rashba \emph{Fiz. Tverd. Tela} (Leningrad)
\textbf{2}, 1224 (1960) [\emph{Sov. Phys. - Solid State} \textbf{2}, 1109
(1960)]; Yu.~A.~Bychkov and E.~I.~Rashba, \emph{J. Phys. C} \textbf{17},
6039 (1984).

\bibitem{DattaDas1990} S. Datta and B. Das, \emph{Appl. Phys. Lett.} \textbf{%
56}, 665 (1990).

\bibitem{Wolf2001} S.~A.~Wolf, D.~D.~Awschalom, R.~A.~Buhrman,
J.~M.~Daughton, S.~von~Moln\`{a}r, M.~L.~Roukes, A.~Y.~Chtchelkanova, and
D.~M.~Treger, \emph{Science} \textbf{294}, 1488 (2001).

\bibitem{lateralinterface} The term ``interface'' throughout the text refers
to the lateral interface between two regions of the two-dimensional electron
gas with different strength of the SO interaction.

\bibitem{Nitta1997} J.~Nitta, T.~Akazaki, H.~Takayanagi, and T.~Enoki, \emph{%
Phys. Rev. Lett.} \textbf{78}, 1335 (1997).

\bibitem{Engels1997} G.~Engels, J.~Lange, Th.~Sch\"{a}pers, and H. L\"{u}th,
\emph{Phys. Rev. B} \textbf{55}, R1958 (1997).

\bibitem{Sato2001} Y.~Sato, T.~Kita, S.~Gozu, and S.~Yamada, \emph{J. Appl.
Phys.} \textbf{89}, 8017 (2001).

\bibitem{Rokhinson2004} L.~P.~Rokhinson, V.~Larkina, Y.~B.~Lyanda-Geller,
L.~N.~Pfeiffer, and K.~W.~West, \emph{Phys. Rev. Lett.} \textbf{93}, 146601
(2004).

\bibitem{Ohio2004} H.~Chen, J.~J.~Heremans, J.~A.~Peters, A.~O.~Govorov,
N.~Goel, S.~J.~Chung, and M.~B.~Santos, \emph{Appl. Phys. Lett.} \textbf{86}%
, 032113 (2005).

\bibitem{Milne} P.~M.~Morse and H.~Feshbach, \emph{Methods of Theoretical
Physics} (McGraw-Hill, New York, 1953).

\bibitem{Nieuwenhuizen} M.~C.~W.~van Rossum and Th.~M.~Niewenhuizen, \emph{%
Rev. Mod. Phys.} \textbf{71}, 313 (1999) and references therein.

\bibitem{valuealpha} We use $\widetilde{\alpha}=0.1$ and $B(x)=const$ in
Fig. \ref{fig:intensity}.

\bibitem{LandauerIBM} R.~Landauer, \emph{IBM J. Res. Dev.} \textbf{1}, 223
(1957); see also \emph{J. Math. Phys.} \textbf{37}, 5259 (1996).

\bibitem{Levinson1989} I.~B.~Levinson, \emph{Zh. Eksp. Teor. Fiz.} \textbf{95%
}, 2175 (1989) [\emph{Sov. Phys. JETP} \textbf{68}, 1257 (1989)].

\bibitem{NozPines} D.~Pines, and P.~Nozieres, \emph{The Theory of Quantum
Liquids} (Addison-Wesley Publishing Co., Inc., Reading, MA, 1989), Vol. I.

\bibitem{Wexler} G.~Wexler, \emph{Proc. Phys. Soc. London} \textbf{89}, 927
(1966).

\bibitem{kineticsPit}
L.~D.~Landau, and E.~M.~Lifshitz, \emph{Course of Theoretical
Physics}, E.~M.~Lifshitz, and L.~P.~Pitaevskii, \emph{Physical
Kinetics}, Vol. 10 (Pergamon Press, Oxford, 1981), Chap. 3, Sec. 37.

\bibitem{Levinson77} I.~B.~Levinson, \emph{Zh. Eksp. Teor. Fiz.} \textbf{73}
, 318 (1977) [\emph{Sov. Phys. JETP} \textbf{46}, 165 (1977)].

\bibitem{DyakonovPerel} M.~I.~Dyakonov, V.~I.~Perel, \emph{Zh. Eksp. Teor.
Fiz.} \textbf{60}, 1954 (1971) [\emph{Sov. Phys. JETP} \textbf{33}, 1053
(1971)].
\end{thebibliography}
\end{document}